\documentclass[aps,prl,twocolumn,groupedaddress]{revtex4-1}

\newcommand{\matJ}{\mathbf{J}}
\newcommand{\Jij}{J_{ij}}
\newcommand{\piex}{\pi_\mathrm{ex}}
\newcommand{\piinh}{\pi_\mathrm{inh}}
\newcommand{\Jex}{J_\mathrm{ex}}
\newcommand{\Jinh}{J_\mathrm{inh}}
\newcommand{\sigex}{\sigma_\mathrm{ex}^2}
\newcommand{\siginh}{\sigma_\mathrm{inh}^2}
\newcommand{\muex}{\mu_\mathrm{ex}}
\newcommand{\muinh}{\mu_\mathrm{inh}}
\newcommand{\zt}{|z|^2}

\makeatletter
\DeclareFontFamily{OMX}{MnSymbolE}{}
\DeclareSymbolFont{MnLargeSymbols}{OMX}{MnSymbolE}{m}{n}
\SetSymbolFont{MnLargeSymbols}{bold}{OMX}{MnSymbolE}{b}{n}
\DeclareFontShape{OMX}{MnSymbolE}{m}{n}{
    <-6>  MnSymbolE5
   <6-7>  MnSymbolE6
   <7-8>  MnSymbolE7
   <8-9>  MnSymbolE8
   <9-10> MnSymbolE9
  <10-12> MnSymbolE10
  <12->   MnSymbolE12
}{}
\DeclareFontShape{OMX}{MnSymbolE}{b}{n}{
    <-6>  MnSymbolE-Bold5
   <6-7>  MnSymbolE-Bold6
   <7-8>  MnSymbolE-Bold7
   <8-9>  MnSymbolE-Bold8
   <9-10> MnSymbolE-Bold9
  <10-12> MnSymbolE-Bold10
  <12->   MnSymbolE-Bold12
}{}
\let\llangle\@undefined
\let\rrangle\@undefined
\DeclareMathDelimiter{\llangle}{\mathopen}%
                     {MnLargeSymbols}{'164}{MnLargeSymbols}{'164}
\DeclareMathDelimiter{\rrangle}{\mathclose}%
                     {MnLargeSymbols}{'171}{MnLargeSymbols}{'171}
\makeatother

\usepackage{graphicx}
\usepackage{amsmath}

\usepackage{tikz}
\usetikzlibrary{external}
\begin{document}

\title{Spectrum density of large sparse random matrices associated to neural
networks}

\author{Herv\'e Rouault}
\email[]{rouaulth@janelia.hhmi.org}
\affiliation{Janelia Research Campus}

\author{Shaul Druckmann}
\email[]{druckmanns@janelia.hhmi.org}
\affiliation{Janelia Research Campus}

\date{\today}

\begin{abstract}
The eigendecomposition of the coupling matrix of large biological networks is
central to the study of the dynamics of these networks.
For neural networks, this matrix should reflect the topology of the network
and conform with Dale's law which states that a neuron can have only all
excitatory or only all inhibitory output connections, i.e., coefficients of one
column of the coupling matrix must all have the same sign.
The eigenspectrum density has been determined before for dense matrices
$J_{ij}$~\cite{rajanabbott}, when several
populations are considered~\cite{PhysRevE.91.012820,wei2012eigenvalue}.
However, the expressions were derived under the assumption of dense
connectivity, whereas neural circuits have sparse connections.
Here, we followed mean-field approaches~\cite{advani2013statistical} in order to
come up with exact self-consistent expressions for the spectrum density in the
limit of sparse matrices for both symmetric and neural network matrices.
Furthermore we introduced approximations that allow for good numerical
evaluation of the density.
Finally, we studied the phenomenology of localization properties of the
eigenvectors.
\end{abstract}


\maketitle

The dynamics of diverse biological systems, such as neural, ecological or genetic
networks involves an interplay between many individual elements.
Since the precise nature of this coupling is difficult to determine, it is
often useful to consider random coupling.
Specifically, in neuroscience synaptic connections between neurons underlie
the dynamics of these networks, yet despite major
efforts~\cite{denk2012structural}, the pattern of these
connections is largely unknown.
Accordingly, models of these dynamics are often studied under the
assumption of random connectivity.
The stability analysis of random networks occupies a central role in the
study of the dynamical behavior of many classes of neural
networks~\cite{brunel1999fast}.
Studying the case of random connectivity is of further importance since it
will serve as an important baseline for deciphering the effects of
more specific connectivity patterns, such as structural
motifs~\cite{milo2002network,song2005highly}.

Like most biological systems, realistic neuronal networks do not have
all-to-all connectivity.
Instead, connectivity is typically highly sparse, i.e., most of the coefficients
of the connectivity matrix $\Jij$ are zero.
The spectrum density of $J_{ij}$ has been previously determined for dense,
$J_{ij}$~\cite{sommers1988spectrum,rajanabbott,PhysRevE.91.012820,
wei2012eigenvalue}.
Here we study the case of highly sparse matrices where the number of non-zero
elements per column is finite.
We are able to derive expressions for the eigenspectrum of sparse networks
obeying the central demarcating line in terms of connectivity structure
in neural circuits, known as Dale's law, that states that neural circuits
are split into two populations: excitatory neurons whose activity
evokes activity in their downstream neurons,
and inhibitory neurons, whose activity suppresses activity
in down stream neurons.
We find striking differences both with the sparse symmetric case where
a non-finite support tail is observed~\cite{semerjian2002sparse,
kuhn2008spectra} and with the dense non-symmetric case respecting Dale's law in
the bulk of the spectrum~\cite{rajanabbott}.

The spectrum of large, sparse, but symmetric random matrices has previously
been studied by several methods~\cite{semerjian2002sparse}.
However, in most biological systems the coupling between units is
non-symmetric.
Here, we develop an approximation scheme based on the cavity method and apply
it to studying the eigenvalue spectrum of non-symmetric matrices (the
application of the method to symmetric matrices is outlined
in~\cite{supplinfo}) and formulated in a different way by
\cite{slanina2011equivalence}.
Next we extend our results to networks whose structure is non-uniform, and
depends on the functional class of a unit, since this is the typical case in
biological networks.
Specifically, in neural circuits, the central demarcating line in terms of
connectivity is that known as Dale's law, splitting neurons into two
populations: excitatory neurons whose activity evokes activity in their
downstream neurons, and inhibitory neurons, whose activity suppresses activity
in down stream neurons.
Accordingly, we develop our methods below to be able to deal with networks
composed of two neural populations.

For general non-symmetric $\Jij$, we follow the field-theoretic mapping of the
eigen-spectrum density of~\cite{feinberg1997non,fyodorov1997almost} and have:
\begin{equation}
  \rho(z) = \frac{1}{\pi N} \partial \partial^* \log \det \begin{pmatrix}
    zI-\Lambda & 0 \\
    0          & z^*I - \Lambda^\dagger
  \end{pmatrix}
\end{equation}
where $\Lambda$ is the Schur decomposition of $\matJ$, $z = x + i y$ and
$\partial = (\partial_x - i \partial_y) / 2$,
$\partial^* = (\partial_x - i \partial_y) / 2$.

With the gaussian integral representation of the determinant and the proper
change of basis, it follows (ignoring irrelevant prefactors within the $\log$)
that:
\begin{equation}
  \rho(z) = -\frac{1}{\pi N} \lim_{\kappa, \kappa' \rightarrow 0} \partial
  \partial^* \log \int d\boldsymbol\psi
  \exp\left(-\mathcal{H}\right)
  \label{eq:specdens}
\end{equation}
where:
\begin{equation}
  \mathcal{H} = \boldsymbol\psi^\dagger\begin{pmatrix}
    \kappa I  & i(z I - J) \\
    i(z^* I - J^\dagger) & \kappa' I
  \end{pmatrix}\boldsymbol\psi
\end{equation}
and integration is over the $2N$-dimensional complex field $\boldsymbol\psi$.
Note that $\kappa$ and $\kappa'$ make the integral well defined but are also
introduced for latter convenience.
The cavity method will consist in isolating a small number of fields from the
partition function $Z = \int d\boldsymbol\psi \exp\left(-\mathcal{H}\right)$.
Let us first isolate the fields corresponding to a neuron $k$:
\begin{equation}
  Z = K_k \int d\psi_k d\psi_{N + k} dh_k dh_{N + k}
  \exp\left(-\mathcal{H}_k\right)
\end{equation}
with:
\begin{multline}
  \mathcal{H}_k =  \kappa\, |\psi_k|^2 + \kappa'\, |\psi_{N + k}|^2
  + 2 i \Re(z \psi_{N + k} \psi_k^*)\\
  - 2 i \Re(h_k \psi_k^*) - 2 i \Re(\psi_{N + k} h_{N + k}^*)
  + 1 / 2\, \mathbf{h}_k^\dagger \mathbf{\Sigma}_k^{-1} \mathbf{h}_k
\end{multline}
where we define the fields acting on $\psi_k$ and $\psi_{N + k}$ as:
\begin{equation}
  \mathbf{h} = \begin{pmatrix}h_h\\ h_{N + k}\end{pmatrix},\quad
  h_k = \sum_{l \neq k} J_{kl} \psi_{N + l},\quad
  h_{N + k} = \sum_{l \neq k} J_{lk}^* \psi_l
\end{equation}
Note that the dependence of $\mathcal{H}_k$ in $h_k$ and $h_{N + k}$ is
quadratic since $Z$ is a gaussian integral and that $\mathbf{\Sigma}_k$
remains to be determined.
The variance of the external fields are expressed as:
\begin{subequations}
\begin{align}
  \langle |h_k|^2 \rangle_{\setminus k} &= 2 \sigma_{h_k}^2
  = \sum_{l,l' \neq k}
  J_{kl}J_{kl'}^* \langle\psi_{N + l} \psi_{N + l'}^*\rangle_{\setminus k}\\
  \langle |h_{N + k}|^2 \rangle_{\setminus k} &= 2 \sigma_{h_{N + k}}^2
  = \sum_{l,l' \neq k}
  J_{lk}^*J_{l'k} \langle\psi_{l} \psi_{l'}^*\rangle_{\setminus k}\\
  \langle h_k h_{N + k}^* \rangle_{\setminus k} &= \sum_{l,l' \neq k}
  J_{kl}J_{l'k} \langle\psi_{N + l} \psi_{l'}^*\rangle_{\setminus k}
\end{align}
\label{eq:he2}
\end{subequations}
Here the moments are computed with a hamiltonian lacking the terms containing
$\psi_k$ and $\psi_{N + k}$.
The cavity approximation consists in neglecting the off-diagonal terms of these
expressions:
\begin{subequations}
\begin{align}
  2 \sigma_{h_k}^2 &\approx \sum_{l \neq k}
  |J_{kl}|^2 \langle |\psi_{N + l}|^2 \rangle_{\setminus k}\\
  2 \sigma_{h_{N + k}}^2 &\approx \sum_{l \neq k}
  |J_{lk}|^2 \langle |\psi_{l}|^2 \rangle_{\setminus k},
  \ \langle h_k h_{N + k}^* \rangle_{\setminus k} \approx 0
\end{align}
\label{eq:cavitapprox}
\end{subequations}
This approximation is thus valid when the network is sparsely connected
(different fields are weakly correlated) or when the mean of the coupling
constants is zero.
It is possible following the same scheme but isolating two fields in the
hamiltonian, to go beyond this approximation and capture finite size effects at
the order $1 / N$.
Performing this integral, we obtain:
\begin{equation}
  Z = K_k 4\pi^4 \sigma_{h_k}^2 \sigma_{h_{N + k}}^2
  / \left(|z|^2 + (\kappa + 2 \sigma_{h_k}^2)(\kappa' + 2 \sigma_{h_{N
  + k}}^2)\right)
\end{equation}
This gives the moments:
\begin{subequations}
\begin{align}
\langle |\psi_k|^2 \rangle &= 2 \sigma_{h_{N + k}}^2
                       / (|z|^2 + 4 \sigma_{h_k}^2 \sigma_{h_{N + k}}^2)\\
\langle |\psi_{N + k}|^2 \rangle &= 2 \sigma_{h_{k}}^2
                       / (|z|^2 + 4 \sigma_{h_k}^2 \sigma_{h_{N + k}}^2)
\end{align}
\label{eq:psie2}
\end{subequations}
Before deriving the eigenspectrum we extend the treatment, dealing with
two population coupling matrices.

Taking the average over the quenched variables, we obtain the relationship
(which is asymptotically exact in the sparse limit):
\begin{multline}
  \llangle \, \langle |\psi_k|^2 \rangle \, \rrangle = \\
  \int \  \left\llangle
  \frac{\prod_{l \neq k} d\mathbf{\pi}_{kl}(J_{kl})d\mathbf{\pi}_{lk}(J_{lk})}
  {\frac{|z|^2}
  {\sum_{l \neq k} |J_{lk}|^2 \langle |\psi_{l}|^2\rangle_{\setminus k}}
  + \sum_{l \neq k} |J_{kl}|^2 \langle |\psi_{N + l}|^2\rangle_{\setminus k}}
  \right\rrangle
  \label{eq:genselfcons}
\end{multline}
A similar expression holds for
$\llangle \, \langle |\psi_{N + k}|^2 \rangle \, \rrangle$

We now consider that we have two populations of neurons: $fN$
columns and $(1 - f)N$ columns of the matrix have coefficients
independently distributed with distribution $\pi_\mathrm{ex}(J)$ and
$\pi_\mathrm{inh}(J)$ respectively. For $\psi_i$ corresponding to an
inhibitory neuron $i$, the measure in the previous integral can now be written:
\begin{equation}
  \prod_{l=1}^{fN}d\piinh(J_{kl})\prod_{l=fN + 1}^{N}d\piex(J_{kl})
  \prod_{l=1}^{N}d\piinh(J_{lk})
  \label{eq:genselfcons2pop}
\end{equation}
We have a similar expression for the $\psi_e$, corresponding to an
excitatory neuron.

From the general expression~(\ref{eq:specdens}), we can derive the spectrum
density:
\begin{subequations}
\begin{align}
  \rho(z) &= \frac{i}{\pi N} \partial \sum_k \langle \psi_k \psi_{N + k}^*
  \rangle \\
  &= \frac{1}{\pi N} \partial \sum_k \frac{z}{|z|^2 + 4 \sigma_{h_k}^2(|z|^2)
  \, \sigma_{h_{N + k}}^2(|z|^2)}
\end{align}
\label{eq:specgene}
\end{subequations}
This expression gives the correct result in the dense limit as well, as shown
below.

In the dense limit, the central limit theorem holds for the
expression~(\ref{eq:cavitapprox}).
It is thus expected that the dense limit expression holds as soon as the number
of connection per neuron tends to infinity in the thermodynamic limit
(\textit{ie} the central limit theorem holds).
In the dense limit, the moments are self-averaging and we can derive the
external field variances, which for an inhibitory column $i$ gives:
\begin{equation}
  2 \sigma_{h_i}^2 \rightarrow N (f \sigex \llangle \langle
  |\psi_{N + e}|^2\rangle \rrangle + (1 - f) \siginh \llangle \langle
  |\psi_{N + i}|^2\rangle \rrangle)
  \label{eq:cavitapproxdense}
\end{equation}
We obtain:
\begin{subequations}
\begin{align}
  \sigma_{h_i}^2 &= \sigma_{h_e}^2\\
  \sigex \sigma_{h_{N + i}}^2
  &= \siginh \sigma_{h_{N + e}}^2
\end{align}
\label{eq:cavitapproxdenseselfcons}
\end{subequations}
and the self-consistency relationship~(\ref{eq:psie2}) gives:
\begin{equation}
  \begin{cases}
    r(\zt) = 0, & \text{if } |z|^2 > z_c^2 \\
    2\, r(\zt) = N \siginh - (1 + \alpha) \zt + \sqrt{\Delta}
    & \text{if } \zt < z_c^2
  \end{cases}
\end{equation}
where $\Delta = (N / \siginh - \zt (1 + \alpha))^2 + 4 \alpha \zt (z_c^2
- \zt)$, $r = 4 \sigma_{h_i}^2 \sigma_{h_{N + i}}^2$ and
$\sigex$, $\siginh$ are the variances of $\piex$,
$\piinh$ respectively.
$z_c^2 = N (f \sigex + (1 - f) \siginh)$.

We finally obtain, by plugging this result into equation~(\ref{eq:specgene}):
\begin{equation}
  \rho(z) = \frac{1}{N \pi \sigex} + \frac{(1 - f)(1 - \alpha)(r(\zt)
  - \zt r'(\zt))}{(\zt + r)^2}
  \label{eq:specgenedense}
\end{equation}

It is straightforward to verify that this expression leads to the same density
as derived previously by Rajan and Abbott~\cite{rajanabbott} for the dense
case.

\begin{figure}
\input{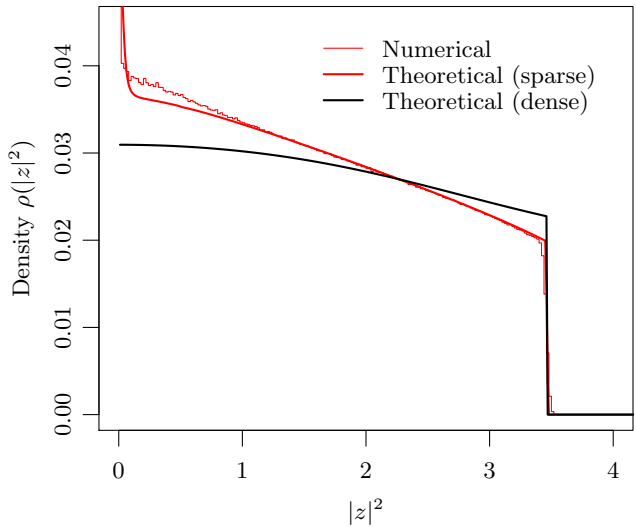}%
\caption{Spectrum density of large sparse random matrices.
  The numerical evaluation comes from a direct eigendecomposition of 4000
  matrices of size 10000x10000.
  The parameters are as follows: $\muex = \muinh = 8$, $J_\mathrm{ex} = 1$,
  $J_\mathrm{inh} = 1.5$ and $f = 0.6$.
  The dense evaluation comes from expression~\ref{eq:specgenedense} whereas the
  sparse evaluation comes from expression~\ref{eq:specgenesparse}.
\label{fig:specdens}}
\end{figure}

In the sparse limit, it is no longer possible to apply the central limit
theorem.
Our approximation entails keeping the previous expression even in the case of
sparse matrices whereby the self-averaging property is not strictly fulfilled.
We begin by considering that every excitatory (resp. inhibitory) neuron has
the same strength of coupling with their postsynaptic partners $J_e$ (resp.
$J_i$) and the noise in the matrix only comes from the presence or absence of
a synaptic connection:
\begin{subequations}
\begin{align}
  \piex(J) &= (1 - \muex / N) \delta(J) + \muex / N\, \delta(J - \Jex)\\
  \piinh(J) &= (1 - \muinh / N) \delta(J) + \muinh / N\, \delta(J + \Jinh)
\end{align}
\end{subequations}
We work in the sparse limit where each neuron has a Poissonian number of
excitatory and inhibitory input neurons.
This leads to the new expression of the self-consistent relationships for
inhibitory neurons $i$ and excitatory neurons $e$:
\begin{equation}
  \llangle \langle |\psi_i|^2 \rangle \rrangle \approx
  \sum_{k,l,m,n = 0}^{+\infty}
  \frac{p(k, l, m, n)\, \Jinh^2 g(k, l)}{|z|^2 + \Jinh^2 g(k, l) h(m, n)}
  \label{eq:genselfconssparse}
\end{equation}
where:
\begin{align*}
  p(k, l, m, n) &= p_{(1 - f)\muinh}(k) p_{f \muinh}(l)
                       p_{(1 - f)\muinh}(m) p_{f \muex}(n) \\
  g(k, l) &= k \llangle \langle |\psi_{i}|^2\rangle \rrangle
                 + l \llangle \langle |\psi_{e}|^2\rangle \rrangle \\
  h(m, n) &= m \Jinh^2 \llangle \langle |\psi_{N + i}|^2\rangle \rrangle
                 + n \Jex^2 \llangle \langle |\psi_{N + e}|^2\rangle \rrangle
\end{align*}
We have a similar expression for
$\llangle \langle |\psi_e|^2 \rangle \rrangle$.
In this expression, $p_\mu(n) = \mu^n e^{-\mu} / n!$ represents the mass of the
Poisson distribution of parameter $\mu$.
In order to apply the self-consistent relationship in the sparse case, we
considered that the neurons connected with the neuron that we isolated in the
mean-field approximation are all equivalent.
This is only true when the neurons are almost all interconnected and form
a large cluster.
In other words, the network needs to have reached the percolation threeshold.
We show in the supplementary information~\cite{supplinfo} that this happens for the
two-population networks when the average number of output connection of
a neuron is higher than one.

It is difficult to solve the self-consistent~(\ref{eq:genselfconssparse})
equations explicitly.
However, it is still straightforward to solve them numerically.
We then obtain thanks to~(\ref{eq:specgene}):
\begin{multline}
  \rho(z) \approx \frac{1}{\pi} \partial
  \sum_{k,l,m,n = 0}^{+\infty}
  z \left( \times \frac{(1 - f)\,p(k, l, m, n)}{\zt + \Jinh^2 g(k, l) h(m, n)}
    \right. \\
    \left. + \frac{f\,q(k, l, m, n)}{\zt + \Jex^2 g(k, l) h(m, n)} \right)
  \label{eq:specgenesparse}
\end{multline}
where:
\begin{equation*}
  q(k, l, m, n) = p_{(1 - f) \muex}(k)\,p_{f\muex}(l)\,p_{(1 - f) \muinh}(m)\,
  p_{f\muex}(n)
\end{equation*}
The comparison between this expression, the expected density from the dense
expression, and the density obtained numerically
is presented in figure~\ref{fig:specdens}.

The sparse limit asymptotic correction to the dense expression are derived
in~\cite{supplinfo} close to the critical value of $|z|^2$ when the spectrum
density becomes zero.
We show that the correction at that point are of order $1 / \mu$ where $\mu$ is
the mean number of outgoing connections of a neuron.
We expect that the order of the correction is the same for the whole bulk of
the spectrum, \textit{ie} outside the divergence in $|z|^2 = 0$.

\begin{figure}
\input{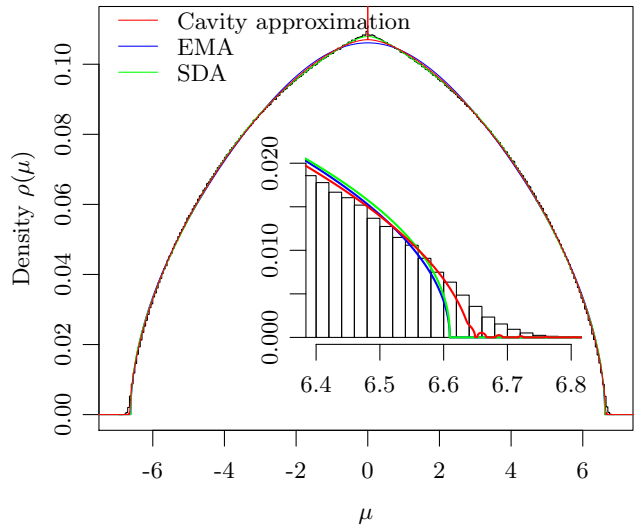}%
\caption{Spectrum density of a sparse symmetric matrix. The different curves
  compare the EMA, SDA and the reported approximation on the tail of the
  spectrum. The average number of connection per neuron is 10. Inset: tail of
  the spectrum.
  \label{fig3}}
\end{figure}

For comparison, we have plotted the spectrum density for symmetric
matrices~\ref{fig3}.
It is known that for symmetric sparse matrices, the tail of the spectrum is
finite contrary to the dense case.
This is well captured by our approximation by contrast with the EMA and SDA.
For non-symmetric matrices, the situation is different and both within our
approximation and by numerical evaluation, we found that the tail of the
spectrum tends to zero with $N$.

It has been shown that for symmetric matrices, the non-finite spectrum support
in the sparse limit is associated with localized eigenvectors on the boundary
of the spectrum.
In the non-symmetric case however, there exists a critical value for the
boundary of the spectrum identical to the one derived from the dense
expression.
We computed the participation ratio of the eigenvector associated with
eigenvalues on the whole spectrum.
This is displayed on figure~\ref{fig2}.
For comparison we plotted the same graph for the dense case in the
supplementary information~\cite{supplinfo}.
We observe that the localization is indeed increased on the boundary of the
spectrum, contrary to the dense case.
However, the localization is weak and tends to zero with the size of the
system (contrary to the symmetric case).

\begin{figure}
\input{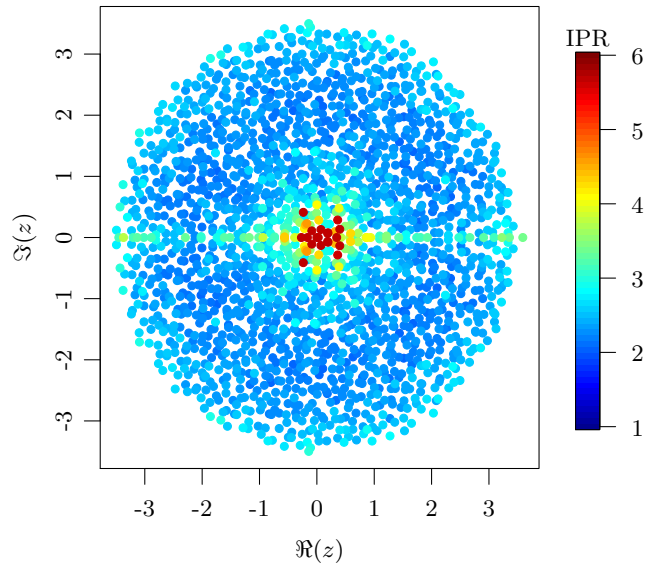}%
\caption{Eigenvalue distribution and the complex plane and inverse
  participation ratio (IPR) of the associated eigenvectors which represent the
  level of sparsity of the eigenvector coefficients associated to the displayed
  eigenvalue.
The corresponding matrix has the same parameters as the one used for
figure~\ref{fig:specdens}.
  \label{fig2}}
\end{figure}


It has been proposed in several works~\cite{legenstein2007edge} that
neural networks stand at the edge of chaos, close to the critical value derived
here.
We have shown that the more realistic assumption of sparse connections
modifies the structure of eigenvalues and eigenvectors around the transition,
and hence the states that appear close to the transition.
Moreover, the influence of sparsity on the tail of the spectrum (which
control the stability behavior of the system) is very different for symmetric
and non-symmetric matrices.
This is due to the lack of feedback on more connected nodes for non-symmetric
matrices.
Our work hence suggests that structural motifs within neural networks (which
include over-represented reciprocal connections) would have a qualitative
influence on the tail of the eigenvalue spectrum of the coupling matrix of
these networks.
This paves the way to studying the non-linear dynamical behavior close to the
transition in the sparse regime, and understanding the relationship between
circuit motif structure and dynamics.

\begin{acknowledgments}
  We would like to thank L. Abbott, D. Hansel, S. Romani and V. Samalam for
  useful comments on the manuscript.
\end{acknowledgments}

\bibliography{refs}

\begin{thebibliography}{16}%
\makeatletter
\providecommand \@ifxundefined [1]{%
 \@ifx{#1\undefined}
}%
\providecommand \@ifnum [1]{%
 \ifnum #1\expandafter \@firstoftwo
 \else \expandafter \@secondoftwo
 \fi
}%
\providecommand \@ifx [1]{%
 \ifx #1\expandafter \@firstoftwo
 \else \expandafter \@secondoftwo
 \fi
}%
\providecommand \natexlab [1]{#1}%
\providecommand \enquote  [1]{``#1''}%
\providecommand \bibnamefont  [1]{#1}%
\providecommand \bibfnamefont [1]{#1}%
\providecommand \citenamefont [1]{#1}%
\providecommand \href@noop [0]{\@secondoftwo}%
\providecommand \href [0]{\begingroup \@sanitize@url \@href}%
\providecommand \@href[1]{\@@startlink{#1}\@@href}%
\providecommand \@@href[1]{\endgroup#1\@@endlink}%
\providecommand \@sanitize@url [0]{\catcode `\\12\catcode `\$12\catcode
  `\&12\catcode `\#12\catcode `\^12\catcode `\_12\catcode `\%12\relax}%
\providecommand \@@startlink[1]{}%
\providecommand \@@endlink[0]{}%
\providecommand \url  [0]{\begingroup\@sanitize@url \@url }%
\providecommand \@url [1]{\endgroup\@href {#1}{\urlprefix }}%
\providecommand \urlprefix  [0]{URL }%
\providecommand \Eprint [0]{\href }%
\providecommand \doibase [0]{http://dx.doi.org/}%
\providecommand \selectlanguage [0]{\@gobble}%
\providecommand \bibinfo  [0]{\@secondoftwo}%
\providecommand \bibfield  [0]{\@secondoftwo}%
\providecommand \translation [1]{[#1]}%
\providecommand \BibitemOpen [0]{}%
\providecommand \bibitemStop [0]{}%
\providecommand \bibitemNoStop [0]{.\EOS\space}%
\providecommand \EOS [0]{\spacefactor3000\relax}%
\providecommand \BibitemShut  [1]{\csname bibitem#1\endcsname}%
\let\auto@bib@innerbib\@empty
\bibitem [{\citenamefont {Rajan}\ and\ \citenamefont
  {Abbott}(2006)}]{rajanabbott}%
  \BibitemOpen
  \bibfield  {author} {\bibinfo {author} {\bibfnamefont {K.}~\bibnamefont
  {Rajan}}\ and\ \bibinfo {author} {\bibfnamefont {L.~F.}\ \bibnamefont
  {Abbott}},\ }\href {\doibase 10.1103/PhysRevLett.97.188104} {\bibfield
  {journal} {\bibinfo  {journal} {Phys. Rev. Lett.}\ }\textbf {\bibinfo
  {volume} {97}},\ \bibinfo {pages} {188104} (\bibinfo {year}
  {2006})}\BibitemShut {NoStop}%
\bibitem [{\citenamefont {Ahmadian}\ \emph {et~al.}(2015)\citenamefont
  {Ahmadian}, \citenamefont {Fumarola},\ and\ \citenamefont
  {Miller}}]{PhysRevE.91.012820}%
  \BibitemOpen
  \bibfield  {author} {\bibinfo {author} {\bibfnamefont {Y.}~\bibnamefont
  {Ahmadian}}, \bibinfo {author} {\bibfnamefont {F.}~\bibnamefont {Fumarola}},
  \ and\ \bibinfo {author} {\bibfnamefont {K.~D.}\ \bibnamefont {Miller}},\
  }\href {\doibase 10.1103/PhysRevE.91.012820} {\bibfield  {journal} {\bibinfo
  {journal} {Phys. Rev. E}\ }\textbf {\bibinfo {volume} {91}},\ \bibinfo
  {pages} {012820} (\bibinfo {year} {2015})}\BibitemShut {NoStop}%
\bibitem [{\citenamefont {Wei}(2012)}]{wei2012eigenvalue}%
  \BibitemOpen
  \bibfield  {author} {\bibinfo {author} {\bibfnamefont {Y.}~\bibnamefont
  {Wei}},\ }\href@noop {} {\bibfield  {journal} {\bibinfo  {journal} {Physical
  Review E}\ }\textbf {\bibinfo {volume} {85}},\ \bibinfo {pages} {066116}
  (\bibinfo {year} {2012})}\BibitemShut {NoStop}%
\bibitem [{\citenamefont {Advani}\ \emph {et~al.}(2013)\citenamefont {Advani},
  \citenamefont {Lahiri},\ and\ \citenamefont
  {Ganguli}}]{advani2013statistical}%
  \BibitemOpen
  \bibfield  {author} {\bibinfo {author} {\bibfnamefont {M.}~\bibnamefont
  {Advani}}, \bibinfo {author} {\bibfnamefont {S.}~\bibnamefont {Lahiri}}, \
  and\ \bibinfo {author} {\bibfnamefont {S.}~\bibnamefont {Ganguli}},\
  }\href@noop {} {\bibfield  {journal} {\bibinfo  {journal} {Journal of
  Statistical Mechanics: Theory and Experiment}\ }\textbf {\bibinfo {volume}
  {2013}},\ \bibinfo {pages} {P03014} (\bibinfo {year} {2013})}\BibitemShut
  {NoStop}%
\bibitem [{\citenamefont {Denk}\ \emph {et~al.}(2012)\citenamefont {Denk},
  \citenamefont {Briggman},\ and\ \citenamefont
  {Helmstaedter}}]{denk2012structural}%
  \BibitemOpen
  \bibfield  {author} {\bibinfo {author} {\bibfnamefont {W.}~\bibnamefont
  {Denk}}, \bibinfo {author} {\bibfnamefont {K.~L.}\ \bibnamefont {Briggman}},
  \ and\ \bibinfo {author} {\bibfnamefont {M.}~\bibnamefont {Helmstaedter}},\
  }\href@noop {} {\bibfield  {journal} {\bibinfo  {journal} {Nature Reviews
  Neuroscience}\ }\textbf {\bibinfo {volume} {13}},\ \bibinfo {pages} {351}
  (\bibinfo {year} {2012})}\BibitemShut {NoStop}%
\bibitem [{\citenamefont {Brunel}\ and\ \citenamefont
  {Hakim}(1999)}]{brunel1999fast}%
  \BibitemOpen
  \bibfield  {author} {\bibinfo {author} {\bibfnamefont {N.}~\bibnamefont
  {Brunel}}\ and\ \bibinfo {author} {\bibfnamefont {V.}~\bibnamefont {Hakim}},\
  }\href@noop {} {\bibfield  {journal} {\bibinfo  {journal} {Neural
  computation}\ }\textbf {\bibinfo {volume} {11}},\ \bibinfo {pages} {1621}
  (\bibinfo {year} {1999})}\BibitemShut {NoStop}%
\bibitem [{\citenamefont {Milo}\ \emph {et~al.}(2002)\citenamefont {Milo},
  \citenamefont {Shen-Orr}, \citenamefont {Itzkovitz}, \citenamefont {Kashtan},
  \citenamefont {Chklovskii},\ and\ \citenamefont {Alon}}]{milo2002network}%
  \BibitemOpen
  \bibfield  {author} {\bibinfo {author} {\bibfnamefont {R.}~\bibnamefont
  {Milo}}, \bibinfo {author} {\bibfnamefont {S.}~\bibnamefont {Shen-Orr}},
  \bibinfo {author} {\bibfnamefont {S.}~\bibnamefont {Itzkovitz}}, \bibinfo
  {author} {\bibfnamefont {N.}~\bibnamefont {Kashtan}}, \bibinfo {author}
  {\bibfnamefont {D.}~\bibnamefont {Chklovskii}}, \ and\ \bibinfo {author}
  {\bibfnamefont {U.}~\bibnamefont {Alon}},\ }\href@noop {} {\bibfield
  {journal} {\bibinfo  {journal} {Science}\ }\textbf {\bibinfo {volume}
  {298}},\ \bibinfo {pages} {824} (\bibinfo {year} {2002})}\BibitemShut
  {NoStop}%
\bibitem [{\citenamefont {Song}\ \emph {et~al.}(2005)\citenamefont {Song},
  \citenamefont {Sj{\"o}str{\"o}m}, \citenamefont {Reigl}, \citenamefont
  {Nelson},\ and\ \citenamefont {Chklovskii}}]{song2005highly}%
  \BibitemOpen
  \bibfield  {author} {\bibinfo {author} {\bibfnamefont {S.}~\bibnamefont
  {Song}}, \bibinfo {author} {\bibfnamefont {P.~J.}\ \bibnamefont
  {Sj{\"o}str{\"o}m}}, \bibinfo {author} {\bibfnamefont {M.}~\bibnamefont
  {Reigl}}, \bibinfo {author} {\bibfnamefont {S.}~\bibnamefont {Nelson}}, \
  and\ \bibinfo {author} {\bibfnamefont {D.~B.}\ \bibnamefont {Chklovskii}},\
  }\href@noop {} {\bibfield  {journal} {\bibinfo  {journal} {PLoS biology}\
  }\textbf {\bibinfo {volume} {3}},\ \bibinfo {pages} {e68} (\bibinfo {year}
  {2005})}\BibitemShut {NoStop}%
\bibitem [{\citenamefont {Sommers}\ \emph {et~al.}(1988)\citenamefont
  {Sommers}, \citenamefont {Crisanti}, \citenamefont {Sompolinsky},\ and\
  \citenamefont {Stein}}]{sommers1988spectrum}%
  \BibitemOpen
  \bibfield  {author} {\bibinfo {author} {\bibfnamefont {H.}~\bibnamefont
  {Sommers}}, \bibinfo {author} {\bibfnamefont {A.}~\bibnamefont {Crisanti}},
  \bibinfo {author} {\bibfnamefont {H.}~\bibnamefont {Sompolinsky}}, \ and\
  \bibinfo {author} {\bibfnamefont {Y.}~\bibnamefont {Stein}},\ }\href@noop {}
  {\bibfield  {journal} {\bibinfo  {journal} {Physical review letters}\
  }\textbf {\bibinfo {volume} {60}},\ \bibinfo {pages} {1895} (\bibinfo {year}
  {1988})}\BibitemShut {NoStop}%
\bibitem [{\citenamefont {Semerjian}\ and\ \citenamefont
  {Cugliandolo}(2002)}]{semerjian2002sparse}%
  \BibitemOpen
  \bibfield  {author} {\bibinfo {author} {\bibfnamefont {G.}~\bibnamefont
  {Semerjian}}\ and\ \bibinfo {author} {\bibfnamefont {L.~F.}\ \bibnamefont
  {Cugliandolo}},\ }\href@noop {} {\bibfield  {journal} {\bibinfo  {journal}
  {Journal of Physics A: Mathematical and General}\ }\textbf {\bibinfo {volume}
  {35}},\ \bibinfo {pages} {4837} (\bibinfo {year} {2002})}\BibitemShut
  {NoStop}%
\bibitem [{\citenamefont {K{\"u}hn}(2008)}]{kuhn2008spectra}%
  \BibitemOpen
  \bibfield  {author} {\bibinfo {author} {\bibfnamefont {R.}~\bibnamefont
  {K{\"u}hn}},\ }\href@noop {} {\bibfield  {journal} {\bibinfo  {journal}
  {Journal of Physics A: Mathematical and Theoretical}\ }\textbf {\bibinfo
  {volume} {41}},\ \bibinfo {pages} {295002} (\bibinfo {year}
  {2008})}\BibitemShut {NoStop}%
\bibitem [{sup()}]{supplinfo}%
  \BibitemOpen
  \href@noop {} {\bibinfo  {journal} {Supplemental information}\ }\BibitemShut
  {NoStop}%
\bibitem [{\citenamefont {Slanina}(2011)}]{slanina2011equivalence}%
  \BibitemOpen
\bibfield  {journal} {  }\bibfield  {author} {\bibinfo {author} {\bibfnamefont
  {F.}~\bibnamefont {Slanina}},\ }\href@noop {} {\bibfield  {journal} {\bibinfo
   {journal} {Physical Review E}\ }\textbf {\bibinfo {volume} {83}},\ \bibinfo
  {pages} {011118} (\bibinfo {year} {2011})}\BibitemShut {NoStop}%
\bibitem [{\citenamefont {Feinberg}\ and\ \citenamefont
  {Zee}(1997)}]{feinberg1997non}%
  \BibitemOpen
  \bibfield  {author} {\bibinfo {author} {\bibfnamefont {J.}~\bibnamefont
  {Feinberg}}\ and\ \bibinfo {author} {\bibfnamefont {A.}~\bibnamefont {Zee}},\
  }\href@noop {} {\bibfield  {journal} {\bibinfo  {journal} {Nuclear Physics
  B}\ }\textbf {\bibinfo {volume} {504}},\ \bibinfo {pages} {579} (\bibinfo
  {year} {1997})}\BibitemShut {NoStop}%
\bibitem [{\citenamefont {Fyodorov}\ \emph {et~al.}(1997)\citenamefont
  {Fyodorov}, \citenamefont {Khoruzhenko},\ and\ \citenamefont
  {Sommers}}]{fyodorov1997almost}%
  \BibitemOpen
  \bibfield  {author} {\bibinfo {author} {\bibfnamefont {Y.~V.}\ \bibnamefont
  {Fyodorov}}, \bibinfo {author} {\bibfnamefont {B.~A.}\ \bibnamefont
  {Khoruzhenko}}, \ and\ \bibinfo {author} {\bibfnamefont {H.-J.}\ \bibnamefont
  {Sommers}},\ }\href@noop {} {\bibfield  {journal} {\bibinfo  {journal}
  {Physics Letters A}\ }\textbf {\bibinfo {volume} {226}},\ \bibinfo {pages}
  {46} (\bibinfo {year} {1997})}\BibitemShut {NoStop}%
\bibitem [{\citenamefont {Legenstein}\ and\ \citenamefont
  {Maass}(2007)}]{legenstein2007edge}%
  \BibitemOpen
  \bibfield  {author} {\bibinfo {author} {\bibfnamefont {R.}~\bibnamefont
  {Legenstein}}\ and\ \bibinfo {author} {\bibfnamefont {W.}~\bibnamefont
  {Maass}},\ }\href@noop {} {\bibfield  {journal} {\bibinfo  {journal} {Neural
  Networks}\ }\textbf {\bibinfo {volume} {20}},\ \bibinfo {pages} {323}
  (\bibinfo {year} {2007})}\BibitemShut {NoStop}%
\end{thebibliography}%


\begin{thebibliography}{2}%
\makeatletter
\providecommand \@ifxundefined [1]{%
 \@ifx{#1\undefined}
}%
\providecommand \@ifnum [1]{%
 \ifnum #1\expandafter \@firstoftwo
 \else \expandafter \@secondoftwo
 \fi
}%
\providecommand \@ifx [1]{%
 \ifx #1\expandafter \@firstoftwo
 \else \expandafter \@secondoftwo
 \fi
}%
\providecommand \natexlab [1]{#1}%
\providecommand \enquote  [1]{``#1''}%
\providecommand \bibnamefont  [1]{#1}%
\providecommand \bibfnamefont [1]{#1}%
\providecommand \citenamefont [1]{#1}%
\providecommand \href@noop [0]{\@secondoftwo}%
\providecommand \href [0]{\begingroup \@sanitize@url \@href}%
\providecommand \@href[1]{\@@startlink{#1}\@@href}%
\providecommand \@@href[1]{\endgroup#1\@@endlink}%
\providecommand \@sanitize@url [0]{\catcode `\\12\catcode `\$12\catcode
  `\&12\catcode `\#12\catcode `\^12\catcode `\_12\catcode `\%12\relax}%
\providecommand \@@startlink[1]{}%
\providecommand \@@endlink[0]{}%
\providecommand \url  [0]{\begingroup\@sanitize@url \@url }%
\providecommand \@url [1]{\endgroup\@href {#1}{\urlprefix }}%
\providecommand \urlprefix  [0]{URL }%
\providecommand \Eprint [0]{\href }%
\providecommand \doibase [0]{http://dx.doi.org/}%
\providecommand \selectlanguage [0]{\@gobble}%
\providecommand \bibinfo  [0]{\@secondoftwo}%
\providecommand \bibfield  [0]{\@secondoftwo}%
\providecommand \translation [1]{[#1]}%
\providecommand \BibitemOpen [0]{}%
\providecommand \bibitemStop [0]{}%
\providecommand \bibitemNoStop [0]{.\EOS\space}%
\providecommand \EOS [0]{\spacefactor3000\relax}%
\providecommand \BibitemShut  [1]{\csname bibitem#1\endcsname}%
\let\auto@bib@innerbib\@empty
\bibitem [{\citenamefont {Wigner}(1958)}]{wigner1958distribution}%
  \BibitemOpen
  \bibfield  {author} {\bibinfo {author} {\bibfnamefont {E.~P.}\ \bibnamefont
  {Wigner}},\ }\href@noop {} {\bibfield  {journal} {\bibinfo  {journal} {Annals
  of Mathematics}\ ,\ \bibinfo {pages} {325}} (\bibinfo {year}
  {1958})}\BibitemShut {NoStop}%
\bibitem [{\citenamefont {Newman}\ \emph {et~al.}(2001)\citenamefont {Newman},
  \citenamefont {Strogatz},\ and\ \citenamefont {Watts}}]{newman2001random}%
  \BibitemOpen
  \bibfield  {author} {\bibinfo {author} {\bibfnamefont {M.~E.}\ \bibnamefont
  {Newman}}, \bibinfo {author} {\bibfnamefont {S.~H.}\ \bibnamefont
  {Strogatz}}, \ and\ \bibinfo {author} {\bibfnamefont {D.~J.}\ \bibnamefont
  {Watts}},\ }\href@noop {} {\bibfield  {journal} {\bibinfo  {journal}
  {Physical review E}\ }\textbf {\bibinfo {volume} {64}},\ \bibinfo {pages}
  {026118} (\bibinfo {year} {2001})}\BibitemShut {NoStop}%
\end{thebibliography}%

\end{document}


\title{Spectrum density of large sparse random matrices associated to neural
networks \\ --- \\ Supplemental information}

\author{Herv\'e Rouault}
\email[]{rouaulth@janelia.hhmi.org}
\affiliation{Janelia Research Campus}

\author{Shaul Druckmann}
\email[]{druckmanns@janelia.hhmi.org}
\affiliation{Janelia Research Campus}


\date{\today}

\pacs{}
\keywords{random matrix, sparse, Dales' law}

\maketitle

\tableofcontents
\section{Symmetric matrices}

We consider a random symmetric matrix $\matJ$ of dimension $N$ whose
coefficients $J_{ij}, 1 \leq i \leq j \leq N$  are identically and
independently distributed with some distribution
$\mathcal{D}$ of mean $0$ and of standard deviation $\sigma$.

Its spectrum is real with eigenvalues $\lambda_i, 1 \leq i \leq N$ and by
definition its spectrum density as a function of $\mu \in \mathbf{R}$ is:
\begin{equation}
  \rho(\mu) = \frac{1}{N} \sumoN \delta(\mu - \lambda_i)
\end{equation}
By using the following representation of the Dirac delta distribution:
\begin{equation}
  \delta(x) = \frac{1}{\pi} \limeps \Im \frac{1}{x - i \epsilon}
\end{equation}
we obtain the following expression for the spectrum density:
\begin{equation}
  \rho(\mu) = \frac{1}{N \pi} \lim_{\epsilon \rightarrow 0} \Im \sumoN
  \partmu \log (\mu -\lambda_i - i \epsilon)
\end{equation}
\begin{equation}
  \rho(\mu) = -\frac{2}{N \pi} \lim_{\epsilon \rightarrow 0} \Im \sumoN
  \partial_\mu \log \int d\phi\, e^{-i(\mu -\lambda_i - i \epsilon)\phi^2}
\end{equation}

After the proper change of basis, we obtain:
\begin{equation}
  \rho(\mu) = -\frac{2}{N \pi} \lim_{\epsilon \rightarrow 0} \Im\, \partmu
  \log Z(\mu - i \epsilon)
\end{equation}
with:
\begin{equation}
  Z(\mu) = \int \prod_{i = 1} ^ N\, d\phi_i
                       \ \exp\left(i\,\mathcal{H}(\bphi)\right)
\end{equation}
and
\begin{equation}
  \mathcal{H}(\boldsymbol\phi) = -\mu\ \bphi^T \bphi + \bphi^T \matJ
  \bphi
\end{equation}

This defines a statistics for the fields $\phi$:
\begin{equation}
  \langle\bullet\rangle = \frac{1}{Z} \int \prod_{i = 1} ^ N\, d\phi_i
  \ \bullet\ \exp\left(i\,\mathcal{H}(\bphi)\right)
\end{equation}

We now exclude the $k$th field from the statistics and define:
\begin{equation}
  \langle\bullet\rangle\mk = \frac{1}{Z\mk} \int \prod_{i \neq k}^N\, d\phi_i\
  \bullet\ \exp\left(i \mathcal{H}\mk(\bphi\mk)\right)
\end{equation}
where $Z\mk$, $\bphi\mk$ and $\mathcal{H}\mk$
are defined like before except the k-th component is removed.

With this definition, we have:
\begin{equation}
  \meanphik = \frac{Z\mk}{Z} \int d\phi_k\,dh_k\ \phi_k^2
  \ e^{-i \mu \phi_k^2 + i h_k \phi_k}
  \left\langle \delta (h_k - \sumkN J_{ki}\phi_i )\right\rangle\mk
\end{equation}

The statistics of the fields being gaussian, $\left\langle \delta\left(h_k
- \sumkN J_{ki}\phi_i\right)\right\rangle\mk$ is gaussian and
we can write:
\begin{equation}
  \meanphik = \frac{1}{Z_k} \int d\phi_k\,dh_k \phi_k^2
  \exp\left(-i\mu\,\phi_k^2 + i h_k \phi_k + i h_k^2 / \sigma_k^2\right)
\end{equation}
where $Z_k$ is the proper normalization constant and:
\begin{equation}
  \sigma_k^2 = -2 i\,\langle h_k^2 \rangle\mk =
  -2 i\,\left\langle \left( \sumkN J_{ki} \phi_i^2 \right)^2
  \right\rangle\mk
\end{equation}
It can be shown that the second moment involving two different fields is
negligible in the the large N limit in the expression (one needs to reproduce
the same calculation as before but excluding fields k and l and computing
$\phi_k\phi_l$).

We have then:
\begin{equation}
  \sigma_k^2 = -2 i\,\sumkN J_{ki}^2 \meanphii\mk
\end{equation}

We can here apply the central limit theorem and obtain:
\begin{equation}
  \sigma_k^2 = -2i\,\sigma^2\sumkN \meanphii\mk
\end{equation}

At the same time, we can easily derive $\meanphik$ as a function of $\mu$ and
$\sigma_k$:
\begin{equation}
  \meanphik = i\frac{d}{d\mu} \log Z_k
\end{equation}
And performing the gaussian integral, we obtain:
\begin{equation}
  Z_k = \frac{\pi}{\sqrt{\mu/\sigma_k^2 + 1 / 4}}
\end{equation}

This gives:
\begin{equation}
  \meanphik = \frac{-i}{2 \mu + \sigma_k^2 / 2}
\end{equation}

By plugging the previously computed $\sigma_k^2$, we obtain:
\begin{equation}
  \meanphik = \frac{-i}{2 \mu - i \sigma^2 \sumkN \meanphii\mk}
\end{equation}

Considering now that $\phi_i^2$ is self-averaging, we obtain the
self-consistent equation:
\begin{equation}
  \meanquenchphi_N = \frac{-i}{2 \mu - i \sigma^2 (N - 1) \meanquenchphi_{N
  - 1}}
\end{equation}

This relationship ensures that in the limit $N \rightarrow +\infty$:
\begin{equation}
  \sqrt{N}\meanquenchphi_N (2 \mu / \sqrt{N} - i \sigma^2 \sqrt{N}
  \meanquenchphi) + i = 0
  \label{eq:selfcons}
\end{equation}

\begin{equation}
  \llangle \rho(\mu) \rrangle = \frac{2}{\pi} \lim_{\epsilon \rightarrow 0}
  \Im\, i \meanquenchphi_N
  \label{eq:rhophi}
\end{equation}

\begin{equation}
  \begin{cases}
    \llangle \rho(\mu) \rrangle = 0    &    \text{if }  \mu > \sqrt{N} \sigma \\
    \llangle \rho(\mu) \rrangle = \frac{2}{\pi N \sigma^2} \sqrt{N\sigma^2
    - \mu^2}  & \text{if }  \mu \leq \sqrt{N} \sigma
  \end{cases}
\end{equation}

This is the Wigner semi-circle law~\cite{wigner1958distribution}.

\section{Spectrum density around the critical point}
\label{sec:spectrum_density_around_the_critical_point}

In this section, we perturbatively expand the equations (20) of the main text
around the critical point.
To lighten the notation, the averaging brackets shall be implicit.
The system of equations writes:

\newcommand{\sklmn}{\sum_{k, l, m, n}}
\newcommand{\pklmn}{p(k, l, m, n)}
\newcommand{\qklmn}{q(k, l, m, n)}
\newcommand{\Jipsi}{\Jinh^2\, (k\, \psi_i^2 + l\, \psi_e^2)}
\newcommand{\Jipsit}{\Jinh^2\, (m\, \psit_i^2 + n\, \psit_e^2)}
\newcommand{\Jepsi}{\Jex^2\, (k\, \psi_i^2 + l\, \psi_e^2)}
\newcommand{\Jepsit}{\Jex^2\, (m\, \psit_i^2 + n\, \psit_e^2)}
\newcommand{\epsilont}{\tilde{\epsilon}}
\begin{subequations}
  \begin{align}
    \psi^2_i &= \sklmn \frac{\pklmn \Jipsi}
    {|z|^2 + \Jipsi(m\, \psit_i^2 + n\, \psit_e^2)} \\
    \psit^2_i &= \sklmn \frac{\pklmn \Jipsit}
    {|z|^2 + \Jipsit(k\, \psi_i^2 + l\, \psi_e^2)} \\
    \psi^2_e &= \sklmn \frac{\qklmn \Jepsi}
    {|z|^2 + \Jepsi(m\, \psit_i^2 + n\, \psit_e^2)} \\
    \psit^2_e &= \sklmn \frac{\qklmn \Jepsit}
    {|z|^2 + \Jepsit(k\, \psi_i^2 + l\, \psi_e^2)}
  \end{align}
  \label{eq:selfconsspar}
\end{subequations}
where $\psit^2_e = \psi_{N + e}^2 / \Jex^2$ and $\psit^2_i = \psi_{N + i}^2
/ \Jinh^2$.
At the critical value of $z$, when the spectrum density becomes zero, the
fields variances tends continuously toward zero.
At the zero-th order, we obtain:
\begin{subequations}
  \begin{align}
    \psi^2_i &= \sklmn \pklmn \Jipsi / |z|^2 \\
    \psi^2_e &= \sklmn \qklmn \Jepsi / |z|^2 \\
    \psit^2_i &= \sklmn \pklmn \Jipsit / |z|^2
  \end{align}
\end{subequations}
It gives the critical value for $|z|^2$ with the field variance ratios:
\begin{subequations}
  \begin{align}
    z_c^2 &= J_i^2 \muinh (1 - f) + \muex J_e^2 f \\
    R &= \frac{\psi^2_e}{\psi^2_i} \approx \frac{\muex J_e^2}{\muinh J_i^2} \\
    \tilde{R} &= \frac{\psit^2_e}{\psit^2_i} \approx \frac{J_e^2}{J_i^2} \\
  \end{align}
\end{subequations}
At the first order, one can obtain the spectrum density close to the critical
value of $z$.
We write the variables at the first order:
\begin{align*}
  |z|^2 &= z_c^2 + \epsilon\\
  \psi_i^2 \psit_i^2 &= P \\
  \psi_e^2 / \psi_i^2 &= R_0 + \epsilon_r = \frac{\muex J_e^2}{\muinh J_i^2}
  + \epsilon_R \\
  \psit_e^2 / \psit_i^2 &= \tilde{R}_0 + \epsilon_r = \frac{J_e^2}{J_i^2}
  + \tilde{\epsilon}_R \\
\end{align*}
With these notations, the self-consistent equations~(\ref{eq:selfconsspar})
write:
\begin{subequations}
  \begin{align}
    \muinh^2 \Jinh^4 f \epsilon_R &= \epsilon \muinh \Jinh^2
    + (\muinh z_c^4 + (1 - f) \muinh^2 \Jinh^4 + f \muex^2 \Jex^4) P \\
  \muex \Jinh^4 f \tilde{\epsilon}_R &= \epsilon \Jinh^2
  + (z_c^4 + (1 - f) \muinh \Jinh^4 + f \muex \Jex^4) P \\
  -\muinh^3 \Jinh^8 (1 - f) \epsilon_R &= \epsilon \muex \Jex^2 \muinh \Jinh^4
  + \muex \Jex^4 (\muex z_c^4 + (1 - f) \muinh^2 \Jinh^4 + f \muex^2 \Jex^4) P \\
  -\muinh^2 \Jinh^8 (1 - f) \tilde{\epsilon}_R &= \epsilon \muinh \Jinh^4
  \Jex^2 + \muex \Jex^4 (z_c^4 + (1 - f) \muinh \Jinh^4 + f \muex \Jex^4) P
  \end{align}
\end{subequations}
By solving these equations, we obtain:
\begin{subequations}
  \begin{align}
    P &= - \frac{z_c^2 \muinh \Jinh^4}{((1-f)\muinh^2 \Jinh^4 + \muex^2
f \Jex^4)(z_c^4 + (1 - f) \muinh \Jinh^4 + f \muex \Jex^4)}\ \epsilon  \\
\epsilon_R &= R_0 \frac{z_c^4 (\muex\Jex^2 - \muinh \Jinh^2) + (\Jex^2 - \Jinh^2)
(\muinh^2 (1-f) \Jinh^4 + \muex^2 f \Jex^4)}{(\muinh^2 (1-f) \Jinh^4 + \muex^2
f \Jex^4) (z_c^4 + (1 - f) \muinh \Jinh^4 + f \muex \Jex^4)} \ \epsilon  \\
\tilde{\epsilon}_R &= \tilde{R}_0 \frac{\muex \Jex^2 - \muinh
\Jinh^2}{(1-f)\muinh^2 \Jinh^4 + f \muex^2 \Jex^4} \ \epsilon 
  \end{align}
\end{subequations}

This allows us to find the spectrum density at the critical point:
\begin{multline}
  \rho(z) \approx \frac{1}{\pi} \partial \, \frac{z}{z_c^2} \sum_{k,l,m,n}
  \left(p(k,l,m,n)(1-f)\left(1 - \frac{\epsilon + \Jinh^2 (k + l R_0)(m
  + n \tilde{R}_0)}{z_c^2}\right) \right. \\
    \left.+ q(k,l,m,n)f \left(1 - \frac{\epsilon + \Jex^2 (k + l R_0)(m
    + n \tilde{R}_0)}{z_c^2}\right)\right)
\end{multline}
\begin{align}
    \rho(z) &\approx \frac{z_c^4}{\pi \muinh \Jinh^4} \frac{-P}{\epsilon}  \\
     &\approx \frac{z_c^6}{\pi ((1-f)\muinh^2 \Jinh^4 + \muex^2
f \Jex^4)(z_c^4 + (1 - f) \muinh \Jinh^4 + f \muex \Jex^4)}
\end{align}

The ratio between the dense and the sparse expression writes:
\begin{equation}
\frac{\rho_\mathrm{dense}(z)}{\rho(z)} = 1 + \frac{(1 - f) \muinh \Jinh^4
+ f \muex \Jex^4}{z_c^4}
\end{equation}
This is a correction of order $1 / \mu$ in the sparsity.

\section{Localization for dense matrices}

We plot here the same graph as the one presented in the main article but for
dense matrices:
\begin{figure}[h!]
  \caption{}
  \centering
   \input{spec_loc-2popdense.tex}
\end{figure}

\section{Percolation threshold for the two-populations network}

We follow here the formalism adopted in~\cite{newman2001random}.
$G_{i1}$ and $G_{e1}$ denote the generating functions of the number of edges
going out of respectively the inhibitory and excitatory nodes.
For the inhibitory generating function:
\begin{equation}
  G_{i0}(x) = \sum^{+\infty}_{k = 0} q_{ik}  x^k
\end{equation}
where $q_{ik}$ is the probability distribution of the number of outgoing
synapses of an inhibitory neuron.
In the main text, we use a Poisson distribution:
\begin{equation}
  q_{ik} = p_{\muinh}(k) = \muinh^k e^{-\muinh} / k!
\end{equation}
We now consider the cluster sizes which are the number of nodes (neurons)
connected together by inhibitory or excitatory connections.
More specifically, we choose a connection at random in the network and consider
its output neuron.
We then consider the output neurons of this neuron, then in turn their output
neurons and so on.
The cluster size of the initial connection is defined as the number of neuron
reached this way.
We denote by $r_s$ the probability of a cluster to be of size $s$ and consider
the associated generating function:
\begin{equation}
  H_{1}(x) = \sum^{+\infty}_{s = 0} q_{is}  x^s
\end{equation}
If we ignore the possibility of a giant cluster containing many recurrent
connections, we can assume that the clusters are tree like, in the limit of
large $N$.
It is then possible to obtain a "Dyson-like" self-consistent relationship
between $H_1$, $G_{i0}$ and $G_{e0}$:
\begin{align}
  H_1(x) &= \underbrace{x\,(f\,q_{e0} + (1 - f) q_{i0})}_\text{no outgoing
    edge} + \underbrace{x\,(f\,q_{e1} + (1 - f) q_{i1}) H_1(x)}_\text{one
    outgoing edge} + \\
    &\hspace*{6em} \underbrace{x\,(f\,q_{e2} + (1 - f) q_{i2}) H_1(x)^2}_\text{two
    outgoing edges} +  \dotsb \\
    &= x\,(f\,G_{e0}(H_1(x) + (1 - f) G_{i0}(H_1(x)))
\end{align}
If we now consider the average number of neurons contained in a cluster, we
obtain:
\begin{equation}
  \langle s \rangle = H_1'(1) = \frac{1}{1 -
                               f\,G_{e0}'(1) - (1 - f) G_{i0}'(1)}
\end{equation}
The so-called percolation transition, appears when $\langle s \rangle$
diverges and a giant cluster appears.
This transition occurs when $(f G_{e0}'(1) + (1 - f) G_{i0}'(1)) = 1$, or when
the average number of outgoing connections from a neuron, irrespective of its
inhibitory of excitatory nature is one.

\bibliography{refs}

